 \documentclass[twocolumn,letterpaper,aps,prd,longbibliography,superscriptaddress,nofootinbib,floatfix]{revtex4-1}

\usepackage{graphicx}	
\usepackage{xspace}	
\usepackage{amsmath}
\usepackage{subfigure}
\usepackage{rotating}

\newcommand{\pt}{\mbox{$p_T$}\xspace}

\newcommand{\pp}{\mbox{$p$$+$$p$}\xspace}

\newcommand{\jpsi}{\mbox{$J/\psi$}\xspace}

\newcommand{\Op}{\mbox{$\mathcal{O}$}}
\newcommand{\sqrts}{\mbox{$\sqrt{s}$}\xspace}
\def\kept#1{\left|#1\right>}
\def\func#1{\left(#1\right)}

\begin{document}

\title{Angular decay coefficients of $J/\psi$ mesons at forward rapidity 
from $p$$+$$p$ collisions at $\sqrt{s}=510$~GeV }

\newcommand{\abilene}{Abilene Christian University, Abilene, Texas 79699, USA}
\newcommand{\augie}{Department of Physics, Augustana University, Sioux Falls, South Dakota 57197, USA}
\newcommand{\banaras}{Department of Physics, Banaras Hindu University, Varanasi 221005, India}
\newcommand{\barc}{Bhabha Atomic Research Centre, Bombay 400 085, India}
\newcommand{\baruch}{Baruch College, City University of New York, New York, New York, 10010 USA}
\newcommand{\bnlcoll}{Collider-Accelerator Department, Brookhaven National Laboratory, Upton, New York 11973-5000, USA}
\newcommand{\bnlphys}{Physics Department, Brookhaven National Laboratory, Upton, New York 11973-5000, USA}
\newcommand{\caucr}{University of California-Riverside, Riverside, California 92521, USA}
\newcommand{\charlesczech}{Charles University, Ovocn\'{y} trh 5, Praha 1, 116 36, Prague, Czech Republic}
\newcommand{\chonbuk}{Chonbuk National University, Jeonju, 561-756, Korea}
\newcommand{\cns}{Center for Nuclear Study, Graduate School of Science, University of Tokyo, 7-3-1 Hongo, Bunkyo, Tokyo 113-0033, Japan}
\newcommand{\colorado}{University of Colorado, Boulder, Colorado 80309, USA}
\newcommand{\columbia}{Columbia University, New York, New York 10027 and Nevis Laboratories, Irvington, New York 10533, USA}
\newcommand{\czechtech}{Czech Technical University, Zikova 4, 166 36 Prague 6, Czech Republic}
\newcommand{\debrecen}{Debrecen University, H-4010 Debrecen, Egyetem t{\'e}r 1, Hungary}
\newcommand{\elte}{ELTE, E{\"o}tv{\"o}s Lor{\'a}nd University, H-1117 Budapest, P{\'a}zm{\'a}ny P.~s.~1/A, Hungary}
\newcommand{\eszterhazy}{Eszterh\'azy K\'aroly University, K\'aroly R\'obert Campus, H-3200 Gy\"ngy\"os, M\'atrai \'ut 36, Hungary}
\newcommand{\ewha}{Ewha Womans University, Seoul 120-750, Korea}
\newcommand{\fsu}{Florida State University, Tallahassee, Florida 32306, USA}
\newcommand{\gsu}{Georgia State University, Atlanta, Georgia 30303, USA}
\newcommand{\hanyang}{Hanyang University, Seoul 133-792, Korea}
\newcommand{\hiroshima}{Hiroshima University, Kagamiyama, Higashi-Hiroshima 739-8526, Japan}
\newcommand{\howard}{Department of Physics and Astronomy, Howard University, Washington, DC 20059, USA}
\newcommand{\ihepprot}{IHEP Protvino, State Research Center of Russian Federation, Institute for High Energy Physics, Protvino, 142281, Russia}
\newcommand{\illuiuc}{University of Illinois at Urbana-Champaign, Urbana, Illinois 61801, USA}
\newcommand{\inrras}{Institute for Nuclear Research of the Russian Academy of Sciences, prospekt 60-letiya Oktyabrya 7a, Moscow 117312, Russia}
\newcommand{\instpasczech}{Institute of Physics, Academy of Sciences of the Czech Republic, Na Slovance 2, 182 21 Prague 8, Czech Republic}
\newcommand{\isu}{Iowa State University, Ames, Iowa 50011, USA}
\newcommand{\jaea}{Advanced Science Research Center, Japan Atomic Energy Agency, 2-4 Shirakata Shirane, Tokai-mura, Naka-gun, Ibaraki-ken 319-1195, Japan}
\newcommand{\jyvaskyla}{Helsinki Institute of Physics and University of Jyv{\"a}skyl{\"a}, P.O.Box 35, FI-40014 Jyv{\"a}skyl{\"a}, Finland}
\newcommand{\kek}{KEK, High Energy Accelerator Research Organization, Tsukuba, Ibaraki 305-0801, Japan}
\newcommand{\korea}{Korea University, Seoul, 136-701, Korea}
\newcommand{\kurchatov}{National Research Center ``Kurchatov Institute", Moscow, 123098 Russia}
\newcommand{\kyoto}{Kyoto University, Kyoto 606-8502, Japan}
\newcommand{\lahorelums}{Physics Department, Lahore University of Management Sciences, Lahore 54792, Pakistan}
\newcommand{\lawllnl}{Lawrence Livermore National Laboratory, Livermore, California 94550, USA}
\newcommand{\losalamos}{Los Alamos National Laboratory, Los Alamos, New Mexico 87545, USA}
\newcommand{\lund}{Department of Physics, Lund University, Box 118, SE-221 00 Lund, Sweden}
\newcommand{\maryland}{University of Maryland, College Park, Maryland 20742, USA}
\newcommand{\mass}{Department of Physics, University of Massachusetts, Amherst, Massachusetts 01003-9337, USA}
\newcommand{\michigan}{Department of Physics, University of Michigan, Ann Arbor, Michigan 48109-1040, USA}
\newcommand{\muhlenberg}{Muhlenberg College, Allentown, Pennsylvania 18104-5586, USA}
\newcommand{\myongji}{Myongji University, Yongin, Kyonggido 449-728, Korea}
\newcommand{\nara}{Nara Women's University, Kita-uoya Nishi-machi Nara 630-8506, Japan}
\newcommand{\natmephi}{National Research Nuclear University, MEPhI, Moscow Engineering Physics Institute, Moscow, 115409, Russia}
\newcommand{\newmex}{University of New Mexico, Albuquerque, New Mexico 87131, USA}
\newcommand{\nmsu}{New Mexico State University, Las Cruces, New Mexico 88003, USA}
\newcommand{\ohio}{Department of Physics and Astronomy, Ohio University, Athens, Ohio 45701, USA}
\newcommand{\ornl}{Oak Ridge National Laboratory, Oak Ridge, Tennessee 37831, USA}
\newcommand{\orsay}{IPN-Orsay, Univ.~Paris-Sud, CNRS/IN2P3, Universit\'e Paris-Saclay, BP1, F-91406, Orsay, France}
\newcommand{\peking}{Peking University, Beijing 100871, People's Republic of China}
\newcommand{\pnpi}{PNPI, Petersburg Nuclear Physics Institute, Gatchina, Leningrad region, 188300, Russia}
\newcommand{\riken}{RIKEN Nishina Center for Accelerator-Based Science, Wako, Saitama 351-0198, Japan}
\newcommand{\rikjrbrc}{RIKEN BNL Research Center, Brookhaven National Laboratory, Upton, New York 11973-5000, USA}
\newcommand{\rikkyo}{Physics Department, Rikkyo University, 3-34-1 Nishi-Ikebukuro, Toshima, Tokyo 171-8501, Japan}
\newcommand{\saispbstu}{Saint Petersburg State Polytechnic University, St.~Petersburg, 195251 Russia}
\newcommand{\seoulnat}{Department of Physics and Astronomy, Seoul National University, Seoul 151-742, Korea}
\newcommand{\stonybrkc}{Chemistry Department, Stony Brook University, SUNY, Stony Brook, New York 11794-3400, USA}
\newcommand{\stonycrkp}{Department of Physics and Astronomy, Stony Brook University, SUNY, Stony Brook, New York 11794-3800, USA}
\newcommand{\tenn}{University of Tennessee, Knoxville, Tennessee 37996, USA}
\newcommand{\titech}{Department of Physics, Tokyo Institute of Technology, Oh-okayama, Meguro, Tokyo 152-8551, Japan}
\newcommand{\tsukuba}{Center for Integrated Research in Fundamental Science and Engineering, University of Tsukuba, Tsukuba, Ibaraki 305, Japan}
\newcommand{\vandy}{Vanderbilt University, Nashville, Tennessee 37235, USA}
\newcommand{\weizmann}{Weizmann Institute, Rehovot 76100, Israel}
\newcommand{\wigner}{Institute for Particle and Nuclear Physics, Wigner Research Centre for Physics, Hungarian Academy of Sciences (Wigner RCP, RMKI) H-1525 Budapest 114, POBox 49, Budapest, Hungary}
\newcommand{\yonsei}{Yonsei University, IPAP, Seoul 120-749, Korea}
\newcommand{\zagreb}{Department of Physics, Faculty of Science, University of Zagreb, Bijeni\v{c}ka 32, HR-10002 Zagreb, Croatia}
\affiliation{\abilene}
\affiliation{\augie}
\affiliation{\banaras}
\affiliation{\barc}
\affiliation{\baruch}
\affiliation{\bnlcoll}
\affiliation{\bnlphys}
\affiliation{\caucr}
\affiliation{\charlesczech}
\affiliation{\chonbuk}
\affiliation{\cns}
\affiliation{\colorado}
\affiliation{\columbia}
\affiliation{\czechtech}
\affiliation{\debrecen}
\affiliation{\elte}
\affiliation{\eszterhazy}
\affiliation{\ewha}
\affiliation{\fsu}
\affiliation{\gsu}
\affiliation{\hanyang}
\affiliation{\hiroshima}
\affiliation{\howard}
\affiliation{\ihepprot}
\affiliation{\illuiuc}
\affiliation{\inrras}
\affiliation{\instpasczech}
\affiliation{\isu}
\affiliation{\jaea}
\affiliation{\jyvaskyla}
\affiliation{\kek}
\affiliation{\korea}
\affiliation{\kurchatov}
\affiliation{\kyoto}
\affiliation{\lahorelums}
\affiliation{\lawllnl}
\affiliation{\losalamos}
\affiliation{\lund}
\affiliation{\maryland}
\affiliation{\mass}
\affiliation{\michigan}
\affiliation{\muhlenberg}
\affiliation{\myongji}
\affiliation{\nara}
\affiliation{\natmephi}
\affiliation{\newmex}
\affiliation{\nmsu}
\affiliation{\ohio}
\affiliation{\ornl}
\affiliation{\orsay}
\affiliation{\peking}
\affiliation{\pnpi}
\affiliation{\riken}
\affiliation{\rikjrbrc}
\affiliation{\rikkyo}
\affiliation{\saispbstu}
\affiliation{\seoulnat}
\affiliation{\stonybrkc}
\affiliation{\stonycrkp}
\affiliation{\tenn}
\affiliation{\titech}
\affiliation{\tsukuba}
\affiliation{\vandy}
\affiliation{\weizmann}
\affiliation{\wigner}
\affiliation{\yonsei}
\affiliation{\zagreb}
\author{A.~Adare} \affiliation{\colorado} 
\author{C.~Aidala} \affiliation{\michigan} 
\author{N.N.~Ajitanand} \affiliation{\stonybrkc} 
\author{Y.~Akiba} \email[PHENIX Spokesperson: ]{akiba@rcf.rhic.bnl.gov} \affiliation{\riken} \affiliation{\rikjrbrc} 
\author{R.~Akimoto} \affiliation{\cns} 
\author{M.~Alfred} \affiliation{\howard} 
\author{V.~Andrieux} \affiliation{\michigan} 
\author{K.~Aoki} \affiliation{\kek} \affiliation{\riken} 
\author{N.~Apadula} \affiliation{\isu} \affiliation{\stonycrkp} 
\author{Y.~Aramaki} \affiliation{\riken} 
% ...
\collaboration{PHENIX Collaboration} \noaffiliation

\date{\today}


\begin{abstract}

We report the first measurement of the full angular distribution for 
inclusive $J/\psi\rightarrow\mu^{+}\mu^{-}$ decays in $p$$+$$p$ collisions at 
$\sqrt{s}=510$~GeV. The measurements are made for $J/\psi$ transverse 
momentum $2<p_{T}<10$~GeV/$c$ and rapidity $1.2<y<2.2$ in the Helicity, 
Collins-Soper, and Gottfried-Jackson reference frames.  In all frames the 
polar coefficient $\lambda_{\theta}$ is strongly negative at low $p_{T}$ 
and becomes close to zero at high $p_{T}$, while the azimuthal coefficient 
$\lambda_{\phi}$ is close to zero at low $p_{T}$, and becomes slightly 
negative at higher $p_{T}$. The frame-independent coefficient 
$\tilde{\lambda}$ is strongly negative at all $p_{T}$ in all frames. The 
data are compared to the theoretical predictions provided by 
nonrelativistic quantum chromodynamics models.

\end{abstract}


\maketitle

\section{Introduction} 

Measurements of heavy quark bound states provide a unique opportunity to 
explore basic quantum chromodynamics (QCD).  Because the energy scale of 
the heavy quark mass is larger than the hadronization scale, 
nonrelativistic QCD (NRQCD) techniques can be applied to provide 
theoretical access to hadronization.  Charmonium, the bound state of a 
charm and anti-charm quark, is an especially convenient laboratory as it 
decays with a considerable branching fraction into two leptons. It is 
composed of two moderately heavy quarks, and is more copiously available 
than bottomonium (a bottom and anti-bottom bound state).

The charmonium wave function can be expressed as a combination of 
intermediate state contributions formed during the $c-\bar{c}$ 
hadronization stage. The S-wave charmonium wave function can be calculated 
from an expansion in a series of the charm and anti-charm velocity $\nu$ 
in the charmonium rest frame \cite{Bodwin:1994jh},

\begin{eqnarray}
\label{eq:s-wave}
 \kept{\psi_{Q}} &=& \Op(1)\kept{^3S_1^{(1)}} +
 \Op(\nu)\kept{^3P_J^{(8)}g}\\\nonumber
 &+& \Op\func{\nu^2}\kept{^3S_1^{(8)}gg} +
\Op\func{\nu^2}\kept{^3S_0^{(8)}g} + \cdots,
\end{eqnarray}

\noindent in the spectroscopic notation $^{2S+1}L_J$. The series contains 
color singlet$^{(1)}$ and color octet$^{(8)}$ states. The nonrelativistic 
operators $\Op$ are parametrized from experimental results.

Several models have been proposed for the production of $J/\psi$ mesons, 
each one with a different interpretation of these intermediate states.  
The Color Evaporation Model (CEM)~\cite{Fritzsch:1977ay}, applied only to 
hadronic collisions, assumes that the nonrelativistic amplitude is 
constant from twice the charm quark mass to twice the D meson mass and 
zero elsewhere.  All relativistic diagrams to a fixed order in $\alpha_s$ 
producing a charm and anti-charm quark in the final state are included.  
The original Color-Singlet Model (CSM)~\cite{Baier:1981uk} explicitly 
requires the $c\bar{c}$ pair produced in the hard scattering to be 
on-shell and in the same quantum state as the hadronized \jpsi 
($^{2S+1}L_{J} =$$^3S_{1}$). The nonrelativistic amplitude is taken as the 
real-space \jpsi wave function evaluated at the origin.  Early 
calculations of the CSM at LO in $\alpha_{s}$ under-predicted cross 
sections at CDF~\cite{Abe:1997jz} and PHENIX~\cite{Adare:2006kf}.  Recent 
calculations at next-to-leading order 
(NLO)~\cite{Artoisenet:2009xh,Chang:2009uj} and next-to-next-to-leading 
order (NNLO)~\cite{Artoisenet:2008fc} increase the predicted cross 
section. NRQCD calculations~\cite{Bodwin:1994jh} predict nonnegligible 
contributions from $c\bar{c}$ production in the color-octet configuration, 
leading to a larger cross section and better agreement with data than the 
current CSM calculations.

Several terms in Eq.~\ref{eq:s-wave} produce similar \jpsi cross sections 
and transverse momentum behavior, but can be experimentally distinguished 
because of their different helicities. The angular distribution of spin 
$\frac{1}{2}$ lepton decays from a spin 1 quarkonium state is derived from 
the density matrix $\rho_{m^{\prime}m}$ (where $m^{\prime}$ and $m$ have 
the possible values --1,0,1) of the production process and parity 
conservation 
constraints~\cite{Jacob:1959at, Collins:1977iv, Gottfried:1964nc}.

The elements of the matrix are identified as

\begin{equation}
  \begin{array}{lllr}
    W_L &=& \rho_{00} & \textrm{(longitudinal helicity)} \\
    W_T &=& \rho_{11} - \rho_{-1-1} & \textrm{(transverse helicity)}\\
    W_{\Delta} &=& \frac{1}{\sqrt{2}} \left(\rho_{10} + \rho_{01}
    \right) &
    \textrm{(single spin-flip)} \\
    W_{\Delta\Delta} &=& \rho_{1-1} & \textrm{(double spin-flip)}  
  \end{array}
\label{eq:density_matrix}
\end{equation}

The angular distribution of the positive lepton from the \jpsi decay can 
be written as

\begin{equation}
\begin{split}
& \frac{dN}{d\cos\theta d\phi} \propto  \\
 & 1 + \lambda_{\theta} \cos^{2}\theta +
 \lambda_{\theta\phi} \sin2\theta \cos\phi +
 \lambda_{\phi}\sin^{2}\theta \cos 2\phi 
\end{split}
\label{eq:angular_dist}
\end{equation}

\noindent where,

\begin{eqnarray}
\begin{aligned}
  \lambda_{\theta~} &=& \frac{W_T - W_L}{W_T + W_L} \\\nonumber
  \lambda_{\phi~} &=& \frac{2 W_{\Delta\Delta}}{W_T + W_L}
  \\\nonumber
  \lambda_{\theta\phi} &=& \frac{\sqrt{2}W_{\Delta}}{W_T + W_L}
\end{aligned}
\end{eqnarray}

\noindent which we call the polar ($\lambda_{\theta}$), the azimuthal 
($\lambda_{\phi}$) and the ``mixed'' ($\lambda_{\theta\phi}$) angular 
decay coefficients.

\begin{figure}[thb]
\includegraphics[width=1.0\linewidth]{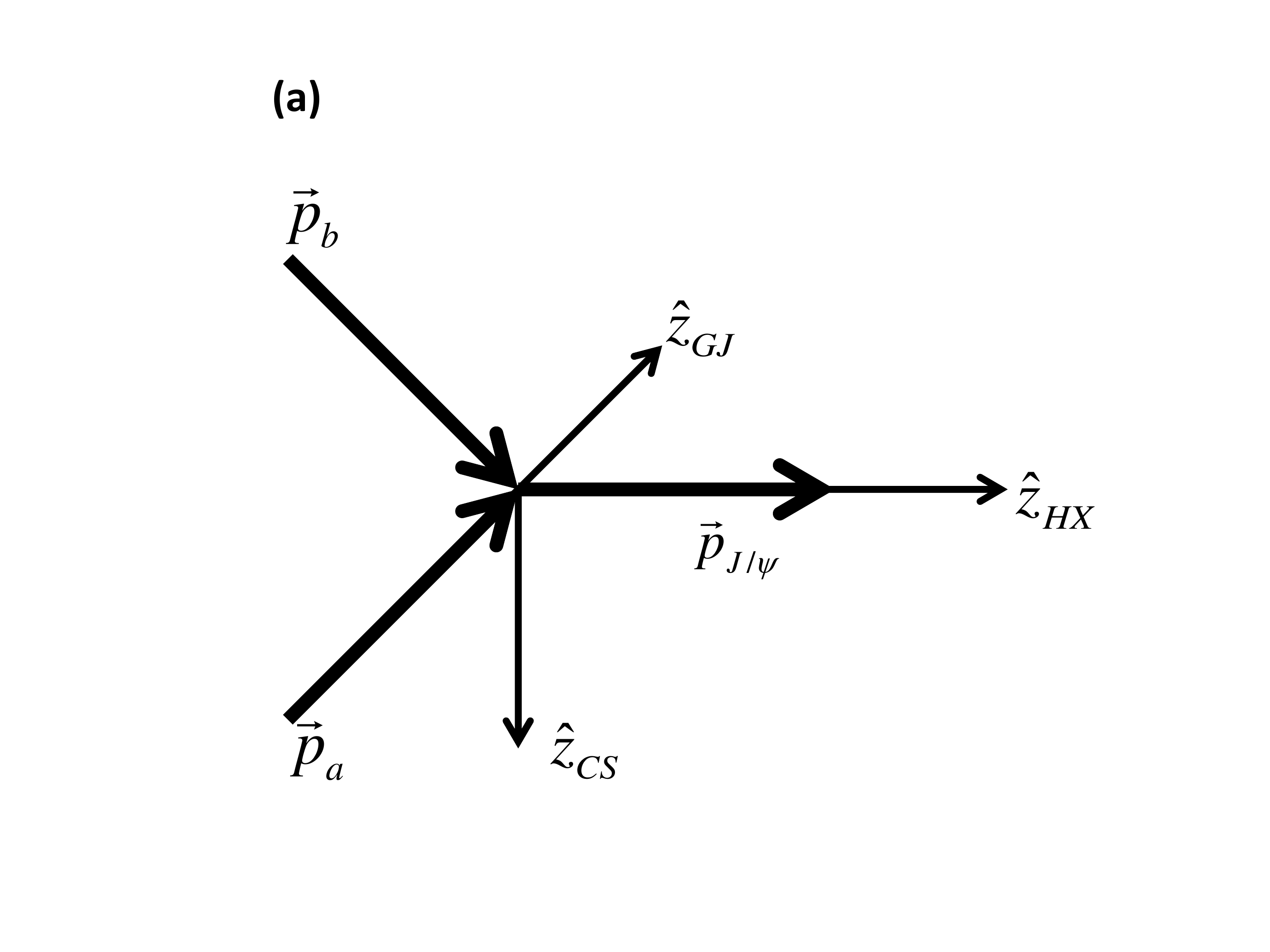}
\includegraphics[width=1.0\linewidth]{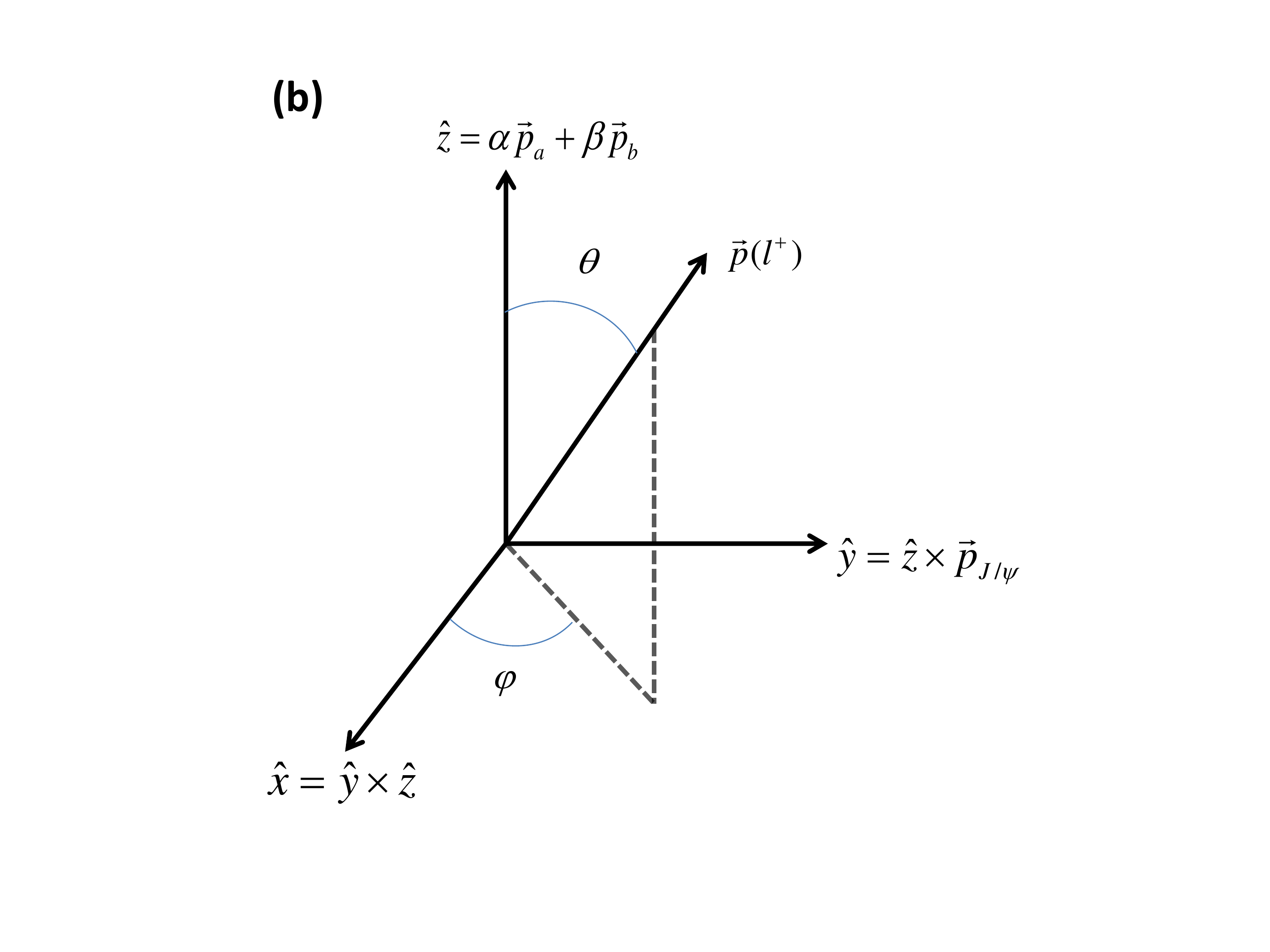}
\caption{\label{fig:ref_frames_coord_sys}
Reference frames and coordinate system used in this analysis. 
The $\hat{x}$ and $\hat{z}$-axes are chosen to lie in the production plane 
determined by the colliding hadrons and the particle produced (a $J/\psi$ 
in this figure). 
(a) shows the production plane and the 
direction convention for the $\hat{z}$ in the Collins-Soper (CS), Helicity 
(HX) and Gottfried-Jackson (GJ) reference frames.
$\vec{p}_a$, $\vec{p}_b$ in this diagram, represent colliding parton momenta.
Note that in an experiment we do not know parton momenta and use colliding
hadron momenta instead.   In (b) the angles $\theta$ and 
$\phi$ represent the direction of the positive 
decay lepton in the corresponding reference frame.
}
\end{figure}

The angles $\phi$ and $\theta$ are measured relative to a reference frame 
defined such that the $\hat{x}$ and $\hat{z}$-axes lie in the production 
plane, formed by the momenta of the colliding protons and the particle 
produced. The direction of the $\hat{z}$-axis within the production plane 
is arbitrary. The simplest frame to study the particle wave function is 
the one in which the density matrix has only diagonal elements, or the 
single and double spin-flip terms are zero. This simplest frame is also 
called the natural frame and is identified when the azimuthal coefficients 
in (\ref{eq:angular_dist}) are zero. The three most common frames used in 
particle angular distribution studies are (Fig. 
\ref{fig:ref_frames_coord_sys}):

\begin{description}

  \item [The Helicity frame (HX)]~\cite{Jacob:1959at}, traditionally used
    in collider experiments, takes the $\hat{z}$-axis as the spin-1 particle
    momentum direction.

    \item [The Collins-Soper frame (CS)]~\cite{Collins:1977iv}, widely
      used in Drell-Yan measurements, chooses the $\hat{z}$-axis as
      the difference between the momenta of the colliding partons
      boosted into the spin-1 particle rest frame.  Note that while the
      original paper~\cite{Collins:1977iv} and subsequent theoretical
      studies used colliding parton momenta in their calculations, 
      the colliding hadron momenta are used here, because we do not have 
      information about the parton momenta.

    \item [The Gottfried-Jackson frame (GJ)]~\cite{Gottfried:1964nc},
      typically used in fixed target experiments, takes the $\hat{z}$-axis 
      as the beam momentum boosted into the spin-1 particle rest frame. At 
      forward angles in a collider environment, the definition of the GJ 
      frame depends heavily on which beam is used in the definition.  If 
      the beam circulating in the same direction as the \jpsi momentum is 
      chosen (GJ forward), the resulting $\hat{z}$-axis is nearly 
      collinear with the $\hat{z}$-axis of the HX and CS frames and points 
      in the same direction. In GJ backward frame (beam circulating in the 
      direction opposite to \jpsi momentum is chosen) the $\hat{z}$-axis 
      points in the opposite direction.

\end{description}

While the angular decay coefficients depend heavily on the reference 
frame, it was noted in~\cite{Faccioli:2008dx} that the $\lambda_{\theta}$ 
coefficient from various measurements transformed into the CS frames 
changes smoothly from longitudinal (negative) to transverse (positive) 
with increasing $J/\psi$ momentum.  The smooth variation occurs between 
measurements from fixed targets by E866/NuSea~\cite{Chang:2003rz} and 
HERA-B~\cite{Abt:2009nu}, as well as a collider environment by 
CDF~\cite{Abulencia:2007us}.  The transformation of the measurements 
depends on the assumption that the $\hat{z}$-axis of the CS frame is the 
natural frame, along which the \jpsi spin-alignment is purely longitudinal 
or transverse.  The assumption is based on measurements of the angular 
distribution for inclusive $J/\psi$ decays from fixed target $p+N$ 
collisions at HERA-B covering $p_T<5$~GeV/$c$ and 
$-0.3<x_F<0.1$~\cite{Abt:2009nu}.  It has been predicted that the natural 
frame at large \pt is near to but not identically along the CS 
$\hat{z}$-axis~\cite{Braaten:2008xg}.  Subsequent work reported 
in~\cite{Faccioli:2010kd} obtained equations which could convert the 
angular parameters measured in one frame to another frame rotated around 
the $\hat{y}$-axis. A combination of polar and azimuthal constants can be 
arranged to form a frame-invariant angular decay coefficient

\begin{equation}
\widetilde{\lambda} = \frac{\lambda_{\theta} + 3\lambda_{\phi}}{1-\lambda_{\phi}}.
\end{equation}

$\widetilde{\lambda}$ is sensitive to the maximum angular asymmetry, or 
polarization, independent of the $\hat{z}$-axis orientation of the 
reference frame. A comparison between $\widetilde{\lambda}$ derived from 
the azimuthal coefficients measured in the different reference frames can 
be used as a consistency check of the parameters extracted from the 
various reference frames.

While there is no clear prediction for the \jpsi spin-alignment from the 
CEM, it has been suggested that multiple soft gluon exchanges destroy the 
spin-alignment of the $c\bar{c}$ pair~\cite{Amundson:1996qr}. Recent 
calculations at NLO~\cite{Artoisenet:2009xh,Chang:2009uj} and 
NNLO~\cite{Artoisenet:2008fc} in the CSM improve agreement with the 
spin-alignment measured previously at PHENIX~\cite{Adare:2009js}, which is 
predicted at NLO to be longitudinal in the HX frame for large 
\pt~\cite{Lansberg:2010vq}. Numerical 
estimates~\cite{Beneke:1996yw,Braaten:1999qk} in the NRQCD approach and 
recent calculations at NLO~\cite{Gong:2008ft} predict a transverse 
spin-alignment in the HX frame at \pt$\gg M_{J/\psi}$ due to gluon 
fragmentation, which disagrees in both sign and magnitude with data from 
CDF~\cite{Abulencia:2007us}.  Measurements of the \jpsi spin alignment in 
different kinematic regions can help distinguish the dominant production 
mechanism.

The PHENIX experiment has already published~\cite{Adare:2009js} a 
$\lambda_{\theta}$ measurement for \jpsi's produced in $p$$+$$p$ 
collisions at $\sqrt{s}=$~200~GeV at midrapidity.  In this paper we 
present a more comprehensive measurement of the full angular distributions 
for the leptonic decays of inclusive \jpsi in $p$$+$$p$ collisions at 
$\sqrt{s}=$~510~GeV for the HX, CS, GJ forward, and GJ backward reference 
frames. The measurement covers a transverse momentum range 
$2<p_{T}<10$~GeV/$c$ and rapidity range $1.2<y<2.2$.

The experimental apparatus used to measure dimuon pairs from \jpsi decays 
is described in Section~\ref{sec:exp}. The procedure followed to obtain 
angular decay coefficients and their uncertainties is explained in 
Section~\ref{sec:analysis}. The results, their comparison to other 
measurements and theoretical predictions are presented in 
Section~\ref{sec:conclusions}.

\section{Experimental Setup and \jpsi Selection}
\label{sec:exp}

The measurements were carried out using the PHENIX 
detector~\cite{Adcox:2003zm} with data from \pp collisions at 
$\sqrt{s}=510$~GeV recorded in 2013.  Decays of 
\jpsi$\rightarrow\mu^{+}\mu^{-}$ were measured in the muon 
spectrometer~\cite{Akikawa:2003zs} for $1.2<y<2.2$ and full azimuthal 
angle.  Collisions are identified by triggering on a minimal multiplicity 
of hits in two beam-beam counters (BBC)~\cite{Allen:2003zt} placed at 
$3.0<|\eta|<3.9$. The data presented correspond to an integrated 
luminosity of 222~pb$^{-1}$.  Approximately $117 \times 10^3 J/\psi$ mesons are 
used to determine the decay coefficients.

The PHENIX muon spectrometer comprises three finely-segmented multi-plane 
cathode strip tracking chambers (MuTr) located in a radial magnetic field 
and positioned in front of five layers of Iarocci tubes interleaved with 
thick steel absorbers (MuID), which provide a hadron rejection of 
$10^{-4}$.  Events containing \jpsi mesons are triggered using logical 
units composed of all tubes in a window projecting from the vertex through 
the MuID.  To satisfy the trigger, trigger logic units in the horizontal 
and vertical projection must contain at least one hit in either the first 
or second layer of the MuID, one additional hit in either the fourth or 
fifth layer, and at least three hits in total.  To avoid the low-momentum 
region where the trigger efficiency changes quickly before reaching a 
plateau, the muons used in this analysis are required to have momentum 
along the beam direction $p_{z}>$1.45~GeV/$c$ as measured at the first 
MuTr station for the spectrometer, corresponding to $\sim$2.1~GeV/$c$ at 
the vertex.

Events are required to occur within 30~cm of the center of the 
experimental apparatus along the beam direction as measured by the 
beam-beam counters.  To improve hadron rejection, a fit of the two tracks 
to the collision vertex was performed and required to have $\chi^{2}<5$ 
per degree of freedom. 
MuTr tracks and MuID hit roads were required to match within four standard deviations to ensure that they correspond to the same particle.

\section{Analysis Procedure}
\label{sec:analysis}

In this section, we outline the procedure used to tune the
simulation to data and extract both the shape of the \jpsi yield and the angular decay coefficients.

\subsection{\jpsi Reconstruction}

\begin{figure}[thb]
\includegraphics[width=0.99\linewidth]{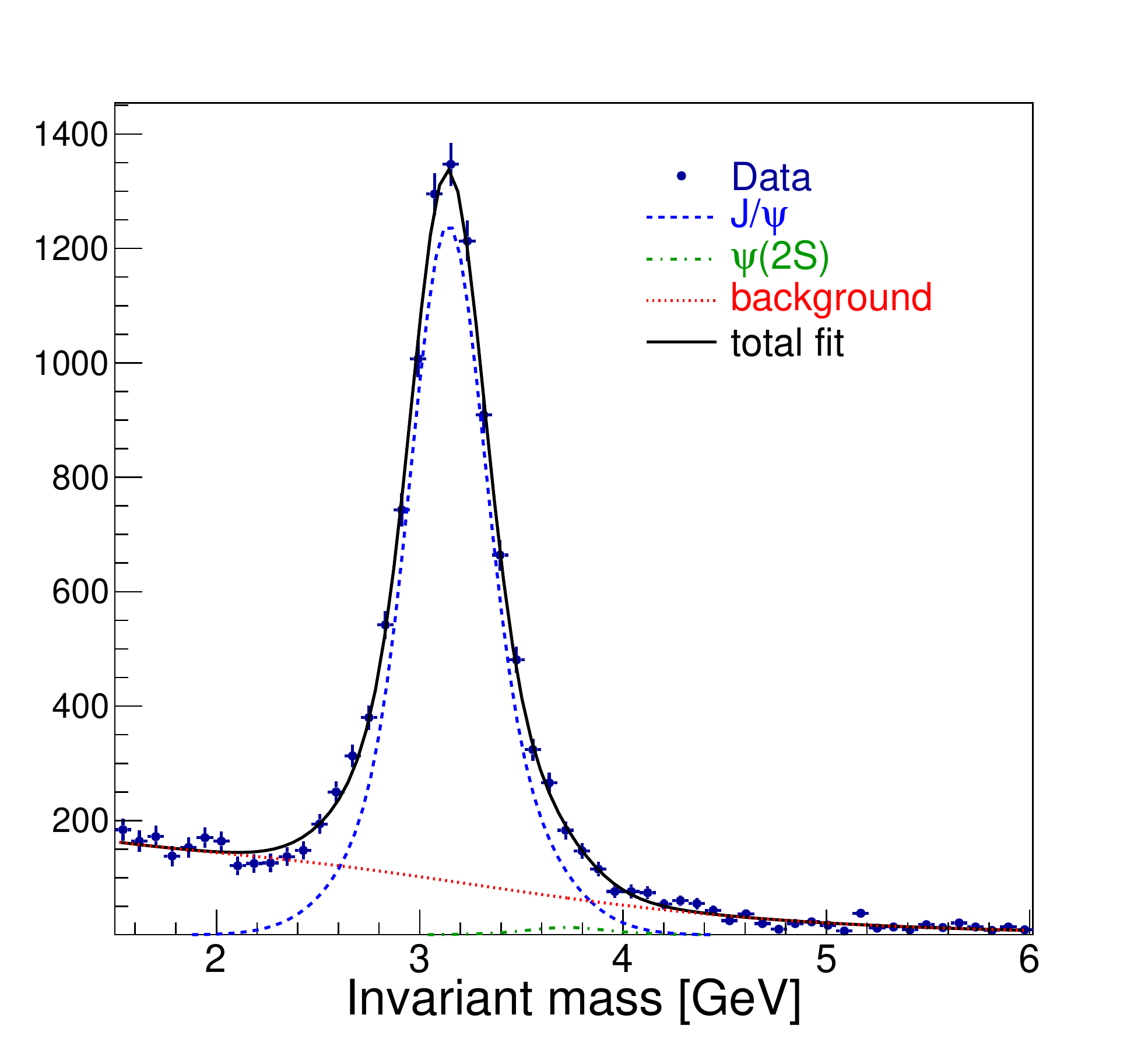}
\caption{\label{fig:yield}
An example of the invariant mass distribution of oppositely charged dimuon 
pairs after like-sign background subtraction (dark blue points) fit with a 
double Gaussian for the \jpsi (dashed blue curve) and $\psi(2S)$ (green 
dash-dotted curve) signals plus exponential for the background (dotted red 
curve). The solid black curve is the sum of signal and background fit.
}
\end{figure}

The \jpsi mesons are reconstructed by calculating the invariant mass of 
all unlike-sign muon pairs after analysis cuts. Combinatorial random 
background is estimated by like-sign dimuons calculated as 
$2\sqrt{N^{++}N^{--}}$, where $N^{++}$ and $N^{--}$ are number of positive 
and negative same-sign pairs respectively, and subtracted. Mass 
distributions for each bin in \pt and rapidity are then fit using a double 
Gaussian as signal and exponential background to remove dimuons from 
Drell-Yan and correlated open-heavy flavor decays (see 
Fig.~\ref{fig:yield}).  The number of \jpsi's is obtained directly by 
integrating the dimuon invariant mass distribution in a mass interval from 
2.5 to 3.7~GeV/$c^{2}$ after background subtraction. Background 
subtraction was performed for each individual $\cos\theta$-$\phi$ bin (see 
Section~\ref{sec:angdec}).

\subsection{Experimental Acceptance and Simulation Tuning}

A simulation of \jpsi mesons generated by tuned {\sc 
pythia}~6.421~\cite{Sjostrand:2000wi} is performed to determine the 
effects of the detector acceptance.  As a complete {\sc 
geant}~3~\cite{GEANT} model of the detector is used to obtain the 
efficiency and acceptance corrections in this analysis, the simulation 
itself needs to be well tuned to reproduce both low-level detector-related 
quantities and high level kinematic distributions. In particular, because 
we perform a two-dimensional fit to the data in $\cos\theta$-$\phi$ space 
for each reference frame, the inefficiencies in the experimental 
acceptance must be properly represented.

To ensure that the acceptance is approximately constant throughout the 
data-taking period, we excluded from analysis the data taken during time 
intervals when the MuTr or MuID had additional tripped high voltage 
channels over normal operation, or there were problems with data 
transmission from the detectors for $>$1\% of all events. Areas of the 
detectors that were disabled or highly inefficient are eliminated in both 
the analyzed data and simulations. In addition, for the MuTr, the charges 
deposited in individual strips within a MuTr cluster are smeared in the 
simulation to match the measured properties in the data.

\begin{figure}[thb]
\includegraphics[width=0.99\linewidth]{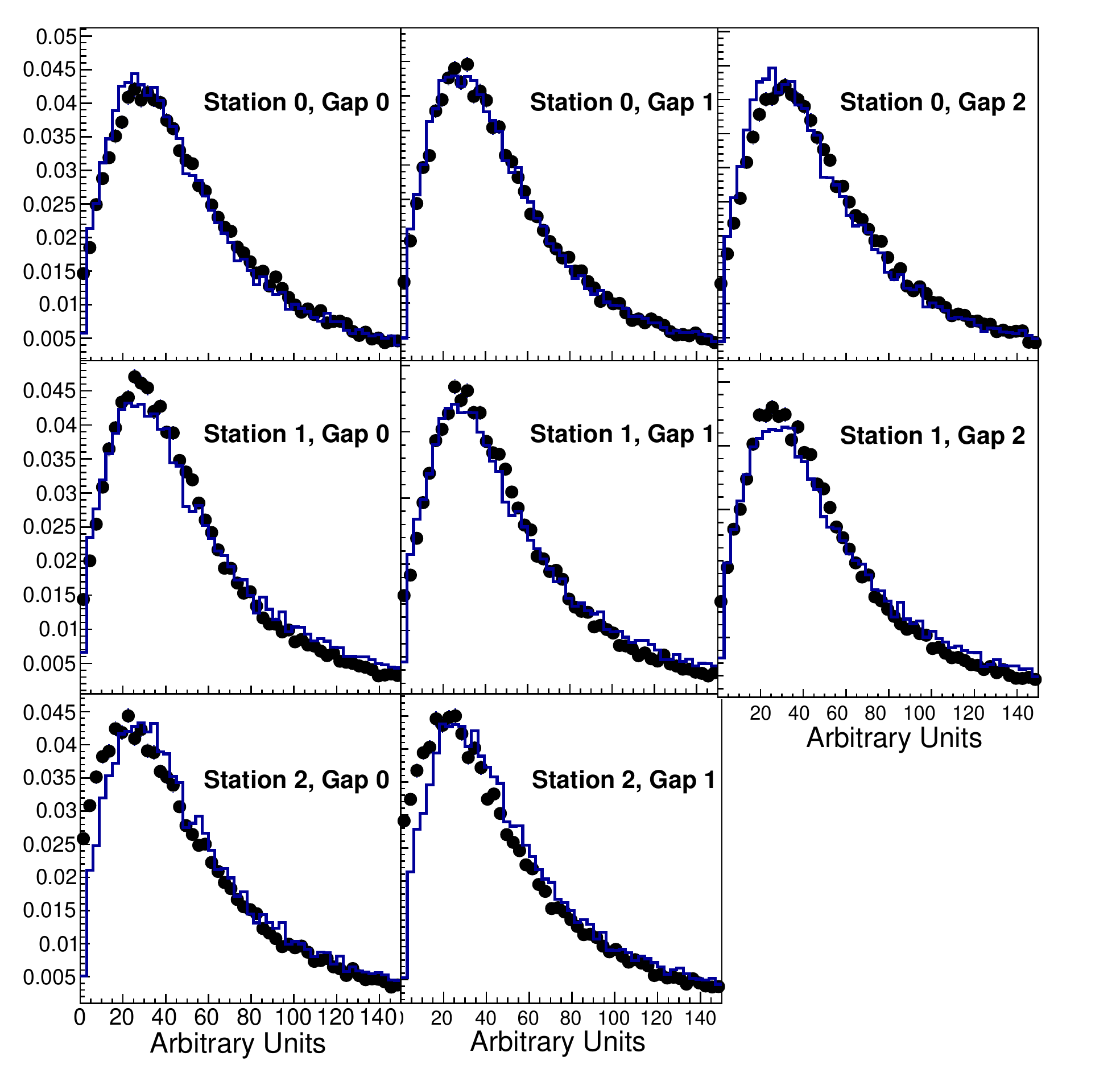}
\caption{\label{fig:cluster_charges}
A comparison of the total cluster charge distributions in the MuTr in 
simulation (blue histogram) and data (closed black circles). Each MuTr 
station is composed of three (Stations 1 and 2) or two (Station 3) 
measurement planes (``gaps'')~\cite{Akikawa:2003zs}. A cluster is the 
collection of ionization energy from the passage of a charged particle in 
the measurement plane.
}
\end{figure}

An example of the excellent agreement between tuned simulations and data 
for the MuTr is shown in Fig.~\ref{fig:cluster_charges}, where cluster 
charge distributions in data and simulation are compared. In addition to 
the low-level performance of the MuTr, the MuID detector has an efficiency 
for pairs of Iarocci tubes that is a function of the collision rate seen 
by the BBC, varying between 0.93 at 400~kHz to 0.88 at 2.2~MHz.  The mean 
efficiency over the course of the running period is used as the efficiency 
of each pair, as a uniform change in efficiency will not affect the 
relative angular acceptance.

\begin{figure}[thb]
\includegraphics[width=0.99\linewidth]{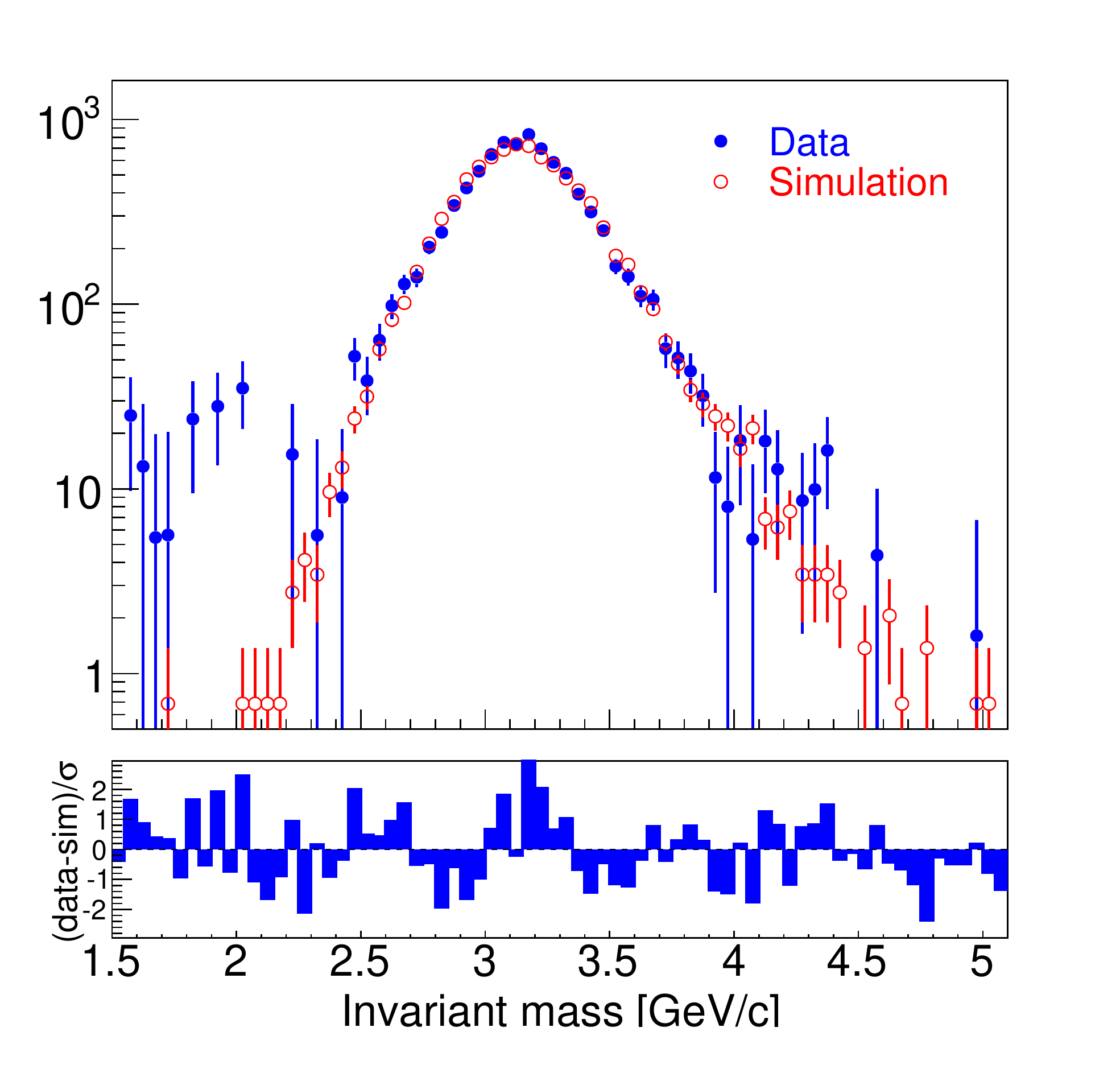}
\caption{\label{fig:invmasscomp}
Invariant mass distribution of simulated \jpsi (red open circles), and 
\jpsi's reconstructed in data (solid blue dots) after all backgrounds are 
subtracted. 
The insert at the bottom shows (data-simulation)/$\sigma$ difference,
where $\sigma$ is the statistical uncertainty of the difference.
}
\end{figure}

At a higher level, a good match of simulation to the data is demonstrated 
in Fig.~\ref{fig:invmasscomp}, where the mass resolution for simulated and 
reconstructed \jpsi's is compared.

Single unpolarized \jpsi's were generated by {\sc 
pythia} and processed through full {\sc geant} 
simulation. Even after the tuning described at the beginning of this chapter, small additional \pt and rapidity weights 
were still required to match the \jpsi's \pt and rapidity distributions in 
{\sc pythia} to those measured experimentally. A systematic uncertainty, 
correlated between data points, was introduced to account for a possible 
mismatch between the \pt and rapidity distributions in simulation and 
data. This systematic uncertainty was estimated by varying the \pt and 
rapidity weights in simulation by 10\%, or one standard deviation of the 
fits to the data (see Section~\ref{sec:syserr} for details). Because the 
detector acceptance in the simulation is sensitive to the input asymmetry 
in the decay muon distributions, the final step in the simulation was to 
apply angular decay coefficients obtained in the initial iteration as 
weights in the simulation, thus imitating the observed \jpsi polarization.

\begin{figure}[thb]
\includegraphics[width=0.99\linewidth]{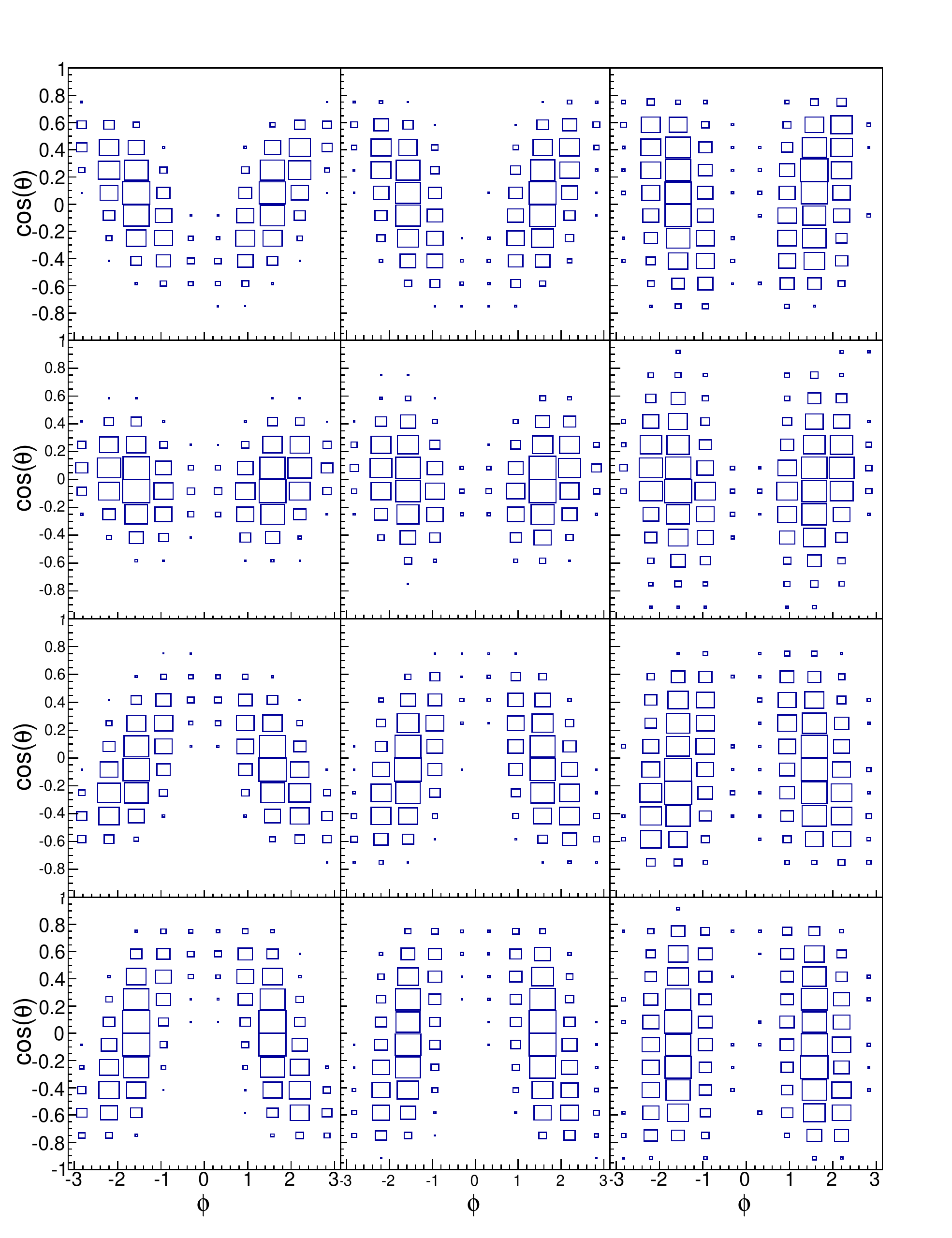}
\caption{\label{fig:thetaphi}
Relative acceptance in $\cos\theta$-$\phi$ bins in (from top to bottom) 
the HX, CS, GJ Backward, and GJ Forward frames for increasing \pt from 
left to right. The area of rectangles is proportional to acceptance value 
in linear scale. See Section~\ref{sec:angdec} for definition of \pt, 
$\cos\theta$, and $\phi$ bins.
}
\end{figure}

The relative acceptance as a function of \pt for the different reference 
frames is shown in Fig.~\ref{fig:thetaphi}.

\subsection{Angular Decay Coefficients}
\label{sec:angdec}

To extract the angular decay coefficients, the background subtracted \jpsi 
yields are histogrammed according to the angular distribution of the 
positive muon in twelve bins of $\cos\theta$ by ten bins of $\phi$, and 
three bins in \pt (2--3, 3--4, and 4--10~GeV$/c$) for each reference frame. 
The mean \pt for each bin are 2.47, 3.46, and 5.45~GeV$/c$, respectively. 
The experimental data are corrected for acceptance, and then fit with 
Eq.~\ref{eq:angular_dist}.  The fit is performed simultaneously in 
$\cos\theta$ and $\phi$ to extract all three angular decay coefficients 
$\lambda_{\theta}$, $\lambda_{\theta\phi}$, $\lambda_{\phi}$, and 
frame-independent $\widetilde{\lambda}$. In general the fits to the data 
are good, with a typical $\chi^2$ value per degree of freedom between 
1.2-2.1, with the number of degrees of freedom typically in the 40--60 
range.

The exact fitting procedure is outlined below.

\begin{enumerate}

\item The $J/\psi$ angular distributions are divided into 12 bins in 
$\cos{\theta}$ and 10 bins in $\phi$. Combinatorial and correlated 
background is subtracted bin-by-bin, and angular distributions are then 
corrected for acceptance, which is calculated assuming no polarization, 
that is $\lambda_{\theta} = \lambda_{\theta\phi} = \lambda_{\phi} = 0$. 
This is done for each of the three transverse momentum bins in each 
polarization frame.

\item $\lambda_{\theta}, \lambda_{\theta\phi}$, and $\lambda_{\phi}$ in 
Eq.~\ref{eq:angular_dist} are varied separately and independently from $-1$ to $1$ with a $0.01$ step, 
and for each step a fit is done to the acceptance corrected measured 
angular distribution. The fit is done for a fixed value of all 
$\lambda$'s. The only free parameter is absolute normalization. A $\chi^2$ 
of the fit is calculated at each step. The minimum $\chi^2$ obtained in 
the three dimensional phase space spanned by $\lambda_\theta$, 
$\lambda_{\theta\phi}$ and $\lambda_\phi$ is chosen as the best fit.

\item Extracted $\lambda$ coefficients are used as weights in the 
simulation to generate acceptance for polarized $J/\psi$ which is used in 
the next iteration. Convergence is achieved when the newly extracted 
$\lambda$ coefficients become zero within the experimental uncertainty, 
which means that the polarization in the simulation matches that in the 
data.

\end{enumerate}

\begin{figure*}[thb]
\begin{minipage}{0.45\linewidth}
\includegraphics[width=0.998\linewidth]{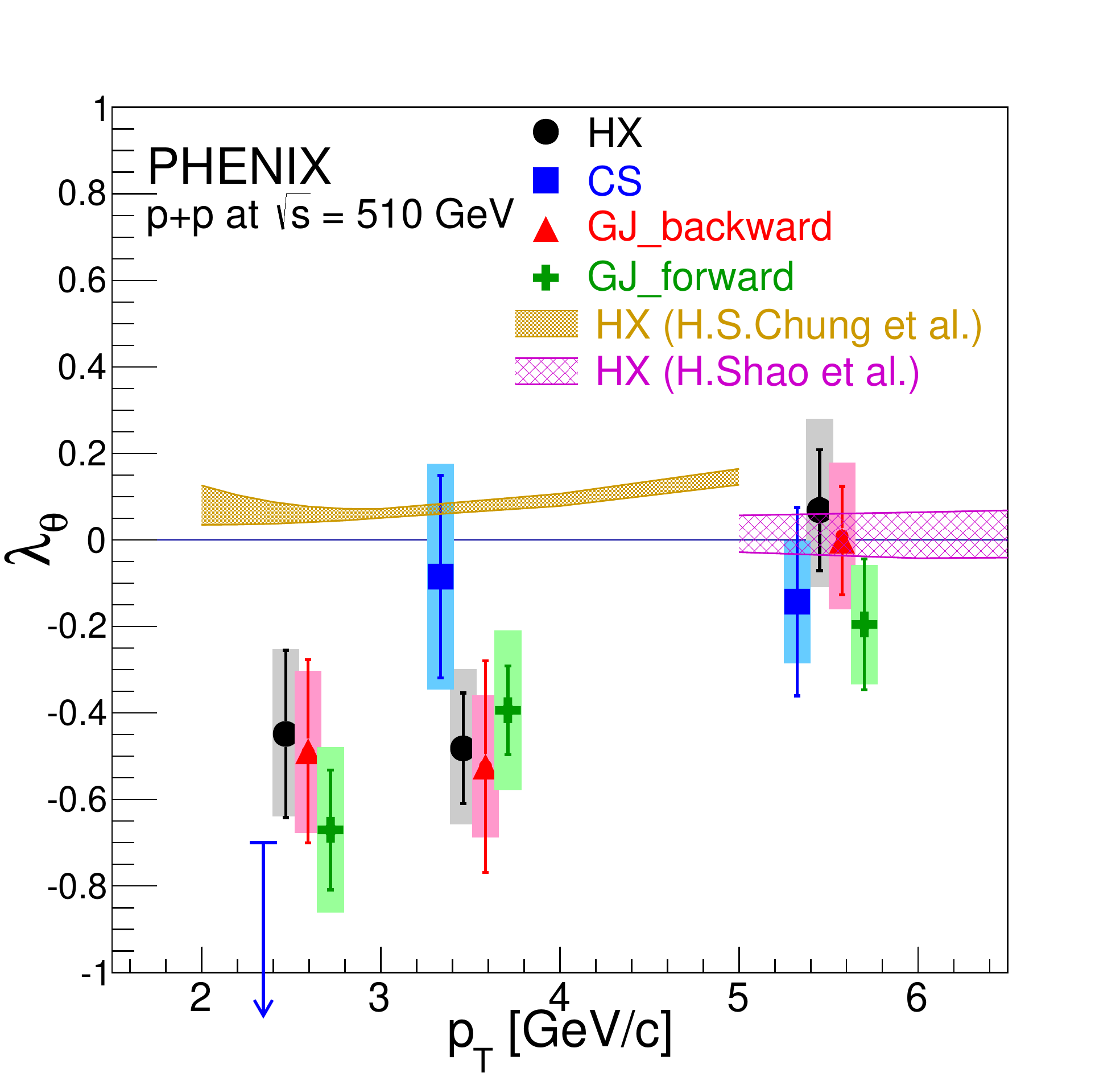}
\caption{\label{fig:finalplot_theta}
Polar angular decay coefficient $\lambda_{\theta}$ as a function of 
transverse momentum for four reference frames and three $p_T$ bins. Black 
circles: HX frame; blue squares: CS frame; red triangles: GJ Backward; 
green crosses: GJ Forward frames.  Shaded error boxes show systematic 
uncertainty.  Points are shifted in \pt for clarity. Down-pointing arrow 
indicates 90\% confidence level upper limit. The data are compared with NRQCD 
theoretical predictions in Helicity frame by H.~S.~Chung et 
al.~\cite{Chung:2010iq} and H.~Shao et al.~\cite{Shao:2014yta}.
}
\end{minipage}
\hspace{0.1cm}
\begin{minipage}{0.45\linewidth}
\includegraphics[width=0.998\linewidth]{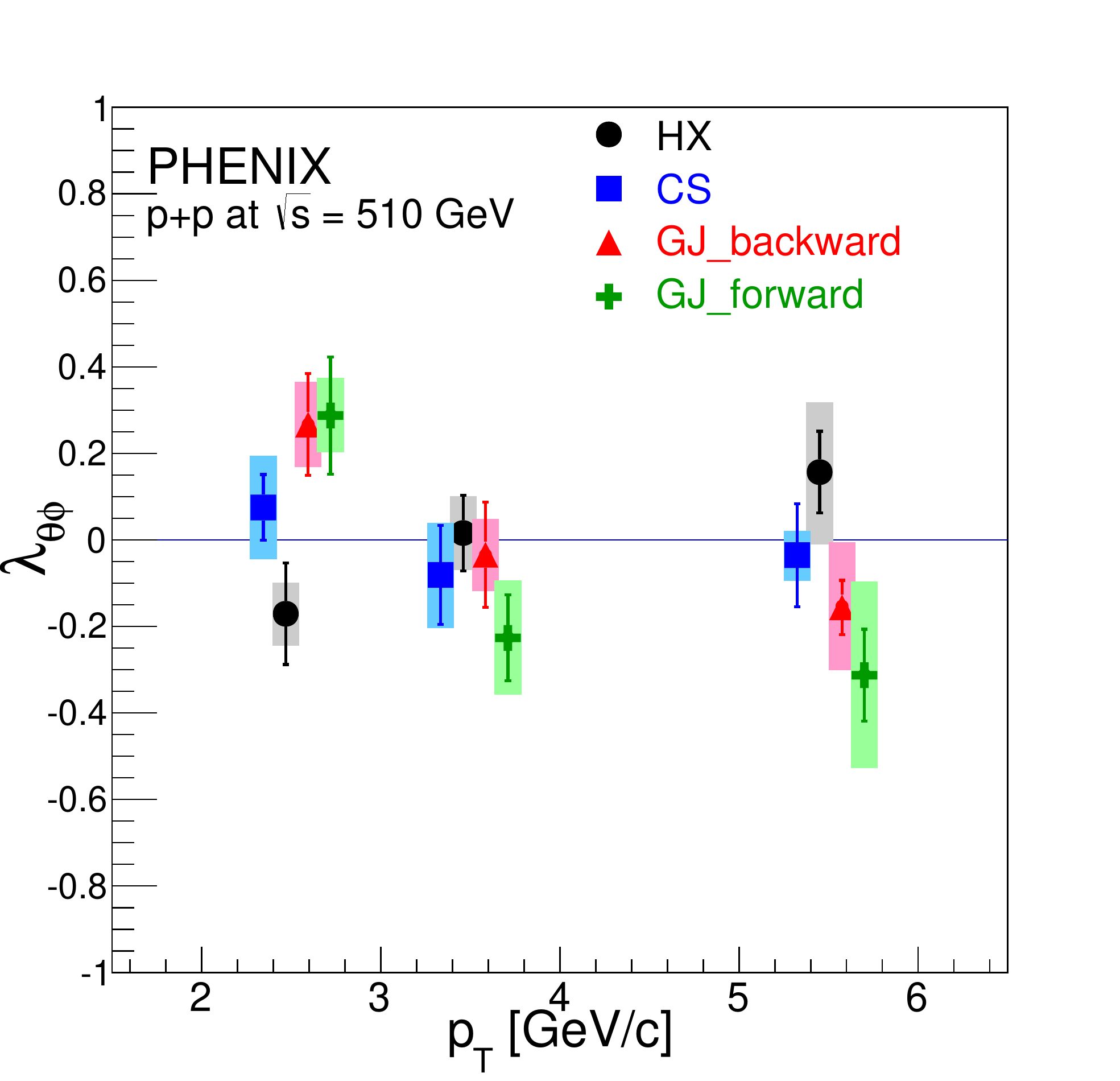}
\caption{\label{fig:finalplot_thetaphi}
``Mixed'' angular decay coefficient $\lambda_{\theta\phi}$ as a function 
of transverse momentum for four reference frames and three $p_T$ bins. 
Black circles: HX frame; blue squares: CS frame; red triangles: GJ 
Backward; green crosses: GJ Forward frames. Shaded error boxes show 
systematic uncertainty.  Points are shifted in \pt for clarity.
}
\end{minipage}
\begin{minipage}{0.45\linewidth}
\includegraphics[width=0.998\linewidth]{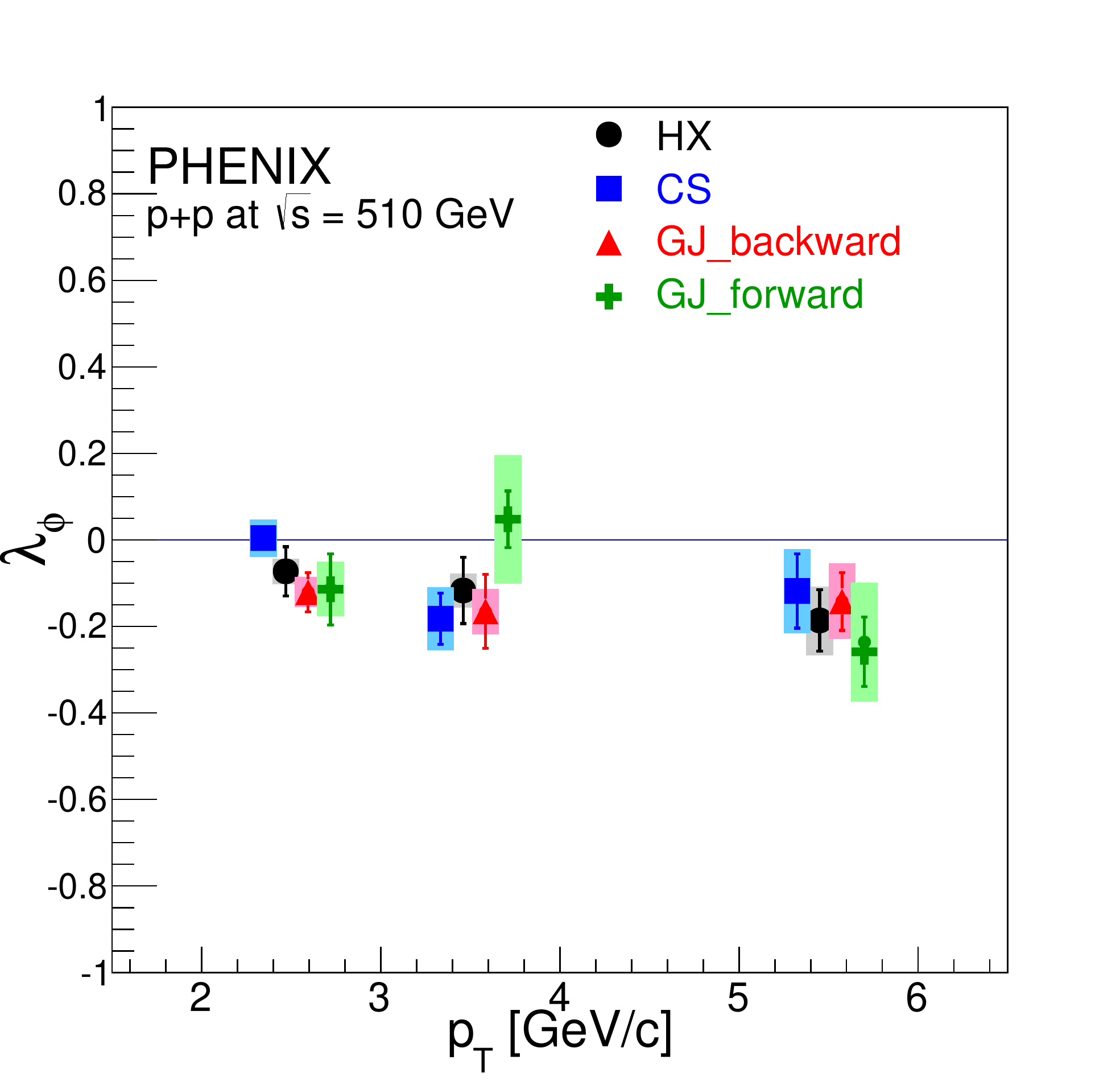}
\caption{\label{fig:finalplot_phi}
Azimuthal angular decay coefficient $\lambda_{\phi}$ as a function of 
transverse momentum for four reference frames and three $p_T$ bins. Black 
circles: HX frame; blue squares: CS frame; red triangles: GJ Backward; 
green crosses: GJ Forward frames. Shaded error boxes show systematic 
uncertainty.  Points are shifted in \pt for clarity.
}
\end{minipage}
\hspace{0.1cm}
\begin{minipage}{0.45\linewidth}
\includegraphics[width=0.998\linewidth]{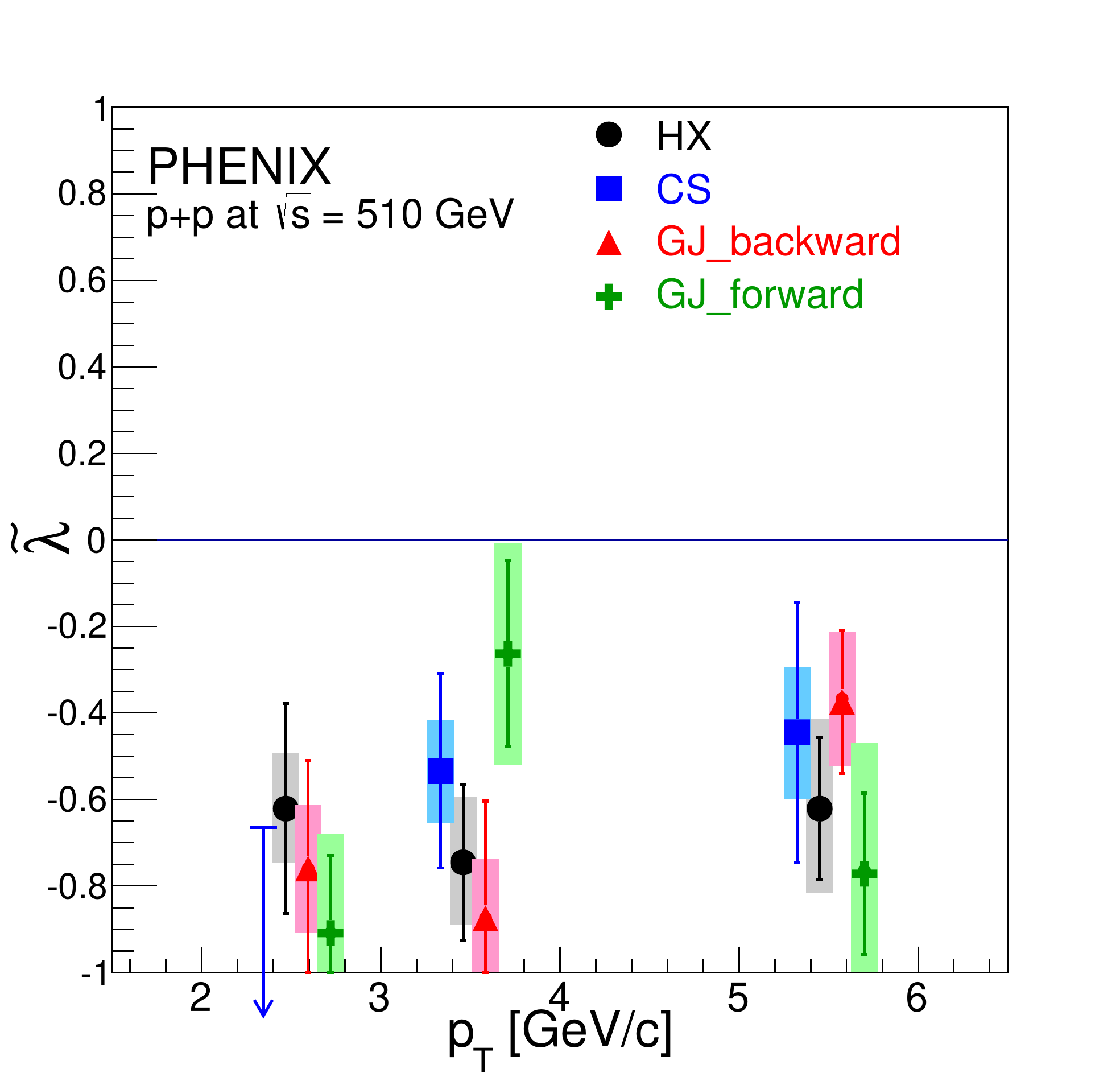}
\caption{\label{fig:finalplot_tilde}
Frame-independent angular decay coefficient $\tilde{\lambda}$ as a 
function of transverse momentum for the four reference frames and three 
$p_T$ bins. Black circles: HX frame; blue squares: CS frame; red 
triangles: GJ Backward; green crosses: GJ Forward frames. Shaded error 
boxes show systematic uncertainty.  Points are shifted in \pt for clarity. 
Down pointing arrow indicates 90\% confidence level upper limit.
}
\end{minipage}
\end{figure*}

The resulting angular decay coefficients $\lambda_{\theta}$, 
$\lambda_{\theta\phi}$, $\lambda_{\phi}$, and frame-independent 
coefficient $\tilde{\lambda}$ are shown in Fig.~\ref{fig:finalplot_theta}, 
Fig.~\ref{fig:finalplot_thetaphi}, Fig.~\ref{fig:finalplot_phi}, and 
Fig.~\ref{fig:finalplot_tilde} respectively, for four reference frames as 
a function of transverse momentum.

\subsection{Systematic Uncertainty Discussion}
\label{sec:syserr}

The statistical uncertainties of the angular decay coefficients were 
calculated by randomizing each bin in $\cos\theta$ vs.~$\phi$ histograms 
with a Gaussian random number according to the statistical uncertainty in 
that bin, and re-fitting. This procedure was repeated one hundred times, 
and the RMS of the resulting $\lambda$ distribution was taken as a 
statistical uncertainty.

A measurement of the angular decay coefficients is sensitive to several 
factors, including the input \pt and rapidity distribution in the 
simulation, exact matching of acceptance between data and simulation, how 
well the simulation reproduces low-level detector-related quantities, and 
time-varying conditions. These uncertainties were estimated by introducing 
variations in the input \pt and rapidity distributions, fiducial cuts, and 
low-level deposited charge smearing in the simulation. Additional 
cross-checks included variations of the collision vertex cut and \jpsi 
rapidity cut.  Possible polarization bias in acceptance was studied with a 
simulation-based blind analysis. In this blind analysis simulated \jpsi's 
generated with a certain polarization were used as fake data. A full 
analysis of the fake data was performed without prior knowledge of the 
input polarization, polarization coefficients were extracted and compared 
to the input values.

The resulting variations in angular decay coefficients were accounted for 
as systematic uncertainties and are listed in 
Table~\ref{tab:typeB_errors}.

\begin{table*}[!ht]
\caption{\label{tab:typeB_errors}
Systematic uncertainties in the  polarization measurement.
}
\begin{ruledtabular}
\begin{tabular}{llcccccccccccc}
      &                         & & $\lambda_{\theta}$ & &  & $\tilde{\lambda}$ & &  & $\lambda_{\phi}$ &   & & $\lambda_{\theta\phi}$ & \\
\hline
 $p_{T}$ bin [GeV/$c$]: & & 2--3  & 3--4   & 4--10   & 2--3    & 3--4   & 4--10  
                         & 2--3  & 3--4   & 4--10   & 2--3    & 3--4   & 4--10  \\
\hline
HX    & Acceptance              & 0.134 & 0.118 & 0.103  & 0.082 & 0.082 & 0.075  & 0.010 & 0.024 & 0.034  & 0.052 & 0.077 & 0.076 \\
      & Kinematics  & $^{+0.049}_{-0.023}$ & $^{+0.050}_{-0.008}$ & $^{+0.120}_{-0.038}$  & $^{+0.036}_{-0.004}$ & $^{+0.042}_{-0.001}$ & $^{+0.089}_{-0.049}$  & $^{+0.006}_{-0.009}$ & $^{+0.002}_{-0.004}$ & $^{+0.021}_{-0.028}$  & $^{+0.003}_{-0.012}$ & $^{+0.014}_{-0.010}$ & $^{+0.012}_{-0.043}$ \\
      & Hit smearing & 0.134 & 0.131 & 0.140   & 0.094 & 0.119 & 0.173  & 0.027 & 0.031 & 0.067  & 0.050 & 0.035 & 0.142 \\
      & Polarization bias       & 0.015 & 0.010  & 0.005       & 0.016 & 0.011    & 0.006  & 0.002 & 0.002  & 0.001  & 0.010 & 0.008  & 0.005 \\
      & TOTAL                   & $^{+0.196}_{-0.191}$ & $^{+0.183}_{-0.176}$ & $^{+0.211}_{-0.178}$      & $^{+0.130}_{-0.125}$ & $^{+0.150}_{-0.144}$ & $^{+0.209}_{-0.195}$  & $^{+0.030}_{-0.031}$ & $^{+0.039}_{-0.039}$ & $^{+0.079}_{-0.081}$  & $^{+0.072}_{-0.073}$ & $^{+0.086}_{-0.085}$ & $^{+0.162}_{-0.167}$ \\
\\
CS    & Acceptance              & 0.106 & 0.148 & 0.079  & 0.076 & 0.101 & 0.083  & 0.010 & 0.010  & 0.025  & 0.061 & 0.076 & 0.042 \\
      & Kinematics  & $^{+0.011}_{-0.004}$ & $^{+0.020}_{-0.014}$ & $^{+0.061}_{-0.066}$    & $^{+0.0147}_{-0.006}$ & $^{+0.014}_{-0.016}$ & $^{+0.068}_{-0.074}$  & $^{+0.003}_{-0.001}$ & $^{+0.001}_{-0.004}$ & $^{+0.020}_{-0.020}$  & $^{+0.016}_{-0.015}$ & $^{+0.011}_{-0.025}$ & $^{+0.019}_{-0.025}$ \\
      & Hit smearing & 0.085 & 0.214 & 0.099   & 0.045 & 0.061 & 0.107  & 0.042 & 0.072 & 0.092  & 0.102 & 0.092 & 0.032 \\
      & Polarization bias       & 0.016 & 0.012  & 0.006       & 0.017 & 0.013 & 0.007  & 0.002 & 0.002  & 0.0015  & 0.011 & 0.009  & 0.007 \\
      & TOTAL                   & $^{+0.136}_{-0.136}$ & $^{+0.261}_{-0.261}$ & $^{+0.141}_{-0.143}$   & $^{+0.089}_{-0.088}$ & $^{+0.118}_{-0.119}$ & $^{+0.151}_{-0.154}$  & $^{+0.043}_{-0.043}$ & $^{+0.073}_{-0.073}$ & $^{+0.098}_{-0.098}$  & $^{+0.120}_{-0.120}$ & $^{+0.120}_{-0.122}$ & $^{+0.057}_{-0.059}$ \\
\\
GJB   & Acceptance              & 0.111 & 0.138 & 0.081    & 0.086 & 0.106 & 0.089  & 0.012 & 0.013  & 0.026  & 0.065 & 0.071 & 0.045 \\
      & Kinematics  & $^{+0.013}_{-0.037}$ & $^{+0.021}_{-0.003}$ & $^{+0.106}_{-0.064}$     & $^{+0.005}_{-0.018}$ & $^{+0.029}_{-0.019}$ & $^{+0.075}_{-0.033}$  & $^{+0.010}_{-0.009}$ & $^{+0.007}_{-0.015}$ & $^{+0.013}_{-0.005}$  & $^{+0.013}_{-0.008}$ & $^{+0.013}_{-0.019}$ & $^{+0.054}_{-0.037}$ \\
      & Hit smearing & 0.149 & 0.087 & 0.121   & 0.119 & 0.082 & 0.112  & 0.032 & 0.050 & 0.083  & 0.074 & 0.041 & 0.133 \\
      & Polarization bias       & 0.018 & 0.009  & 0.004       & 0.019 & 0.010 & 0.005  & 0.002 & 0.002  & 0.002  & 0.016 & 0.009  & 0.005 \\
      & TOTAL                   & $^{+0.186}_{-0.189}$ & $^{+0.165}_{-0.163}$ & $^{+0.180}_{-0.159}$   & $^{+0.147}_{-0.148}$ & $^{+0.137}_{-0.136}$ & $^{+0.162}_{-0.147}$  & $^{+0.035}_{-0.035}$ & $^{+0.052}_{-0.054}$ & $^{+0.088}_{-0.087}$  & $^{+0.099}_{-0.098}$ & $^{+0.083}_{-0.084}$ & $^{+0.151}_{-0.145}$ \\
\\
GJF   & Acceptance              & 0.129 & 0.122 & 0.120  & 0.081 & 0.084 & 0.078  & 0.015 & 0.026  & 0.035  & 0.061 & 0.076 & 0.074 \\
      & Kinematics  & $^{+0.005}_{-0.000}$ & $^{+0.024}_{-0.020}$ & $^{+0.008}_{-0.020}$    & $^{+0.029}_{-0.019}$ & $^{+0.007}_{-0.006}$ & $^{+0.096}_{-0.013}$  & $^{+0.017}_{-0.009}$ & $^{+0.006}_{-0.016}$ & $^{+0.112}_{-0.002}$  & $^{+0.022}_{-0.016}$ & $^{+0.023}_{-0.000}$ & $^{+0.044}_{-0.026}$ \\
      & Hit smearing & 0.141 & 0.137 & 0.067   & 0.212 & 0.243 & 0.276  & 0.060 & 0.145 & 0.110  & 0.058 & 0.106 & 0.200 \\
      & Polarization bias       & 0.015 & 0.012  & 0.006       & 0.016 & 0.013 & 0.007  & 0.002 & 0.001  & 0.001  & 0.013 & 0.010  & 0.007 \\
      & TOTAL                   & $^{+0.192}_{-0.191}$ & $^{+0.185}_{-0.184}$ & $^{+0.137}_{-0.139}$    & $^{+0.229}_{-0.227}$ & $^{+0.257}_{-0.257}$ & $^{+0.303}_{-0.287}$  & $^{+0.064}_{-0.063}$ & $^{+0.148}_{-0.149}$ & $^{+0.160}_{-0.115}$  & $^{+0.087}_{-0.086}$ & $^{+0.133}_{-0.131}$ & $^{+0.217}_{-0.214}$ \\
\end{tabular} \end{ruledtabular}
\end{table*}

\begin{table*}[!ht]
\caption{\label{tab:results1}
$\lambda_{\theta}$ and $\tilde{\lambda}$ in four frames. Only statistical 
errors are shown. Mean \pt for each of the three bins are 2.47, 3.46, and 
5.45~GeV/$c$ respectively.  The numbers in the CS frame for the 
$p_T=2$--3~GeV/$c$ bin are 90\% confidence level upper limits.
}
\begin{ruledtabular}
\begin{tabular}{lcccccc}
                      & &          $\lambda_{\theta}$ &                       &                 & $\tilde{\lambda}$ &  \\
\hline
 $p_{T}$ bin [GeV/$c$]: & 2--3  & 3--4   & 4--10   & 2--3    & 3--4   & 4--10  \\
\hline
 HX   & -0.449 $\pm$ 0.195 & ~~-0.482 $\pm$ 0.131 & ~~~0.069 $\pm$ 0.142    
      & -0.621 $\pm$ 0.241 & ~~-0.745 $\pm$ 0.180 & ~~-0.621 $\pm$ 0.163 \\
 CS   &  \textless -0.701          & ~~-0.085 $\pm$ 0.238 & ~~-0.143 $\pm$ 0.221    
      &  \textless -0.665           & ~~-0.534 $\pm$ 0.221 & ~~-0.445 $\pm$ 0.305 \\
 GJB  & -0.489 $\pm$ 0.218 & ~~-0.524 $\pm$ 0.252 & ~~-0.002 $\pm$ 0.134    
      & -0.760 $\pm$ 0.256 & ~~-0.875 $\pm$ 0.279 & ~~-0.375 $\pm$ 0.171 \\
 GJF  & -0.670 $\pm$ 0.141 & ~~-0.394 $\pm$ 0.105 & ~~-0.195 $\pm$ 0.151    
      & -0.909 $\pm$ 0.185 & ~~-0.263 $\pm$ 0.221 & ~~-0.772 $\pm$ 0.190 \\
\hline
\end{tabular} \end{ruledtabular}
\end{table*}

\begin{table*}[!ht]
\caption{\label{tab:results2}
$\lambda_{\phi}$ and $\lambda_{\theta\phi}$ in four frames. Only 
statistical errors are shown. Mean \pt for each of the three bins are 
2.47, 3.46, and 5.45~GeV/$c$ respectively.
}
\begin{ruledtabular}
\begin{tabular}{lcccccc}
                      & &          $\lambda_{\phi}$ &                       &                 & $\lambda_{\theta\phi}$ &  \\
\hline
 $p_{T}$ bin [GeV/$c$]: & 2--3  & 3--4   & 4--10   & 2--3    & 3--4   & 4--10  \\
\hline
 HX                   & -0.073 $\pm$ 0.057 & ~~-0.117 $\pm$ 0.077 & ~~-0.186 $\pm$ 0.071   & -0.171 $\pm$ 0.120   & ~~~0.016 $\pm$ 0.087 & ~~~0.157 $\pm$ 0.094 \\
 CS                   & ~0.004 $\pm$ 0.027 & ~~-0.182 $\pm$ 0.059 & ~~-0.118 $\pm$ 0.086   & ~0.075 $\pm$ 0.076   & ~~-0.081 $\pm$ 0.110  & ~~-0.035 $\pm$ 0.120 \\
 GJB                  & -0.121 $\pm$ 0.045 & ~~-0.165 $\pm$ 0.085 & ~~-0.142 $\pm$ 0.067   & ~0.267 $\pm$ 0.120   & ~~-0.034 $\pm$ 0.120  & ~~-0.156 $\pm$ 0.063 \\
 GJF                  & -0.114 $\pm$ 0.082 & ~~~0.048 $\pm$ 0.065 & ~~-0.259 $\pm$ 0.080   & ~0.288 $\pm$ 0.140   & ~~-0.230 $\pm$ 0.100  & ~~-0.313 $\pm$ 0.110 \\
\end{tabular} \end{ruledtabular}
\end{table*}

The total systematic uncertainty shown as shaded boxes in 
Figs.~\ref{fig:finalplot_theta} through~\ref{fig:finalplot_tilde} is taken 
to be the quadratic sum of these components, assuming that they are 
uncorrelated.

\section{Results and Discussion}
\label{sec:conclusions}

We have presented the first measurement of the full angular distribution 
from \jpsi decays to muons in $p$$+$$p$ collisions at $\sqrt{s}$~=~510~GeV 
at forward rapidity ($1.2<y<2.2$) in the Helicity, Collins-Soper, and 
Gottfried-Jackson reference frames. The results are summarized in 
Tables~\ref{tab:results1} and~\ref{tab:results2}, and in 
Figs.~\ref{fig:finalplot_theta} through~\ref{fig:finalplot_tilde}.

The measurements presented here are for inclusive \jpsi. Feed-down from 
higher mass quarkonium states also contribute to the observed polarization 
and is not separated out.

In all frames the polar coefficient $\lambda_{\theta}$ is strongly 
negative at low $p_{T}$ and becomes close to zero at high $p_{T}$, while 
the azimuthal coefficient $\lambda_{\phi}$ is close to zero at low 
$p_{T}$, and becomes slightly negative at higher $p_{T}$. The 
frame-independent coefficient $\tilde{\lambda}$ is strongly negative at 
all $p_{T}$ in all frames. Consistency of $\tilde{\lambda}$ values in all 
polarization frames indicates that systematic uncertainties are well under 
control. The obtained polarization coefficient $\tilde{\lambda}$ is in 
good agreement with what was reported by the STAR 
experiment~\cite{Trzeciak:2015xmn}, for the same \sqrts at midrapidity and 
higher transverse momentum.
 
At the Large Hadron Collider (LHC), the LHCb 
experiment~\cite{Aaij:2013nlm} reported similar, although smaller values 
of $\lambda_{\theta}$ with similar trend in transverse momentum at forward 
rapidity. $\lambda_{\theta}$ measured by the ALICE 
experiment~\cite{Abelev:2011md} at forward rapidity is consistent with no 
polarization, although, within experimental uncertainty, it can be said to 
be similar to the LHCb result. A very comprehensive CMS 
measurement~\cite{Chatrchyan:2013cla} indicates that both 
$\lambda_{\theta}$ and $\tilde{\lambda}$ are consistent with zero.  
However, note that the CMS measurement covers much a higher 
transverse momentum range and for more central rapidities.

The measured polar coefficient $\lambda_{\theta}$ is compared to 
theoretical prediction for prompt \jpsi in Helicity frame calculated in 
the NRQCD factorization approach by H.~S.~Chung et al.~\cite{Chung:2010iq} 
and H.~Shao~\cite{Shao:2014yta} in Fig.~\ref{fig:finalplot_theta}. At high 
transverse momentum both predictions are in good agreement with the data, 
while at low $p_{T}$ a strong deviation can be seen. While theory expects 
$\lambda_{\theta}$ to be small and slightly positive at low \pt, it is 
strongly negative in the data. The polar coefficient result in the 
Helicity frame poses a challenge to the NRQCD effective theory at low \pt, 
where perturbative calculations are more difficult to compute. No 
theoretical calculation is available for the frame-independent coefficient 
$\tilde{\lambda}$ or for other reference frames. The reported experimental 
results represent a challenge for the theory and provide a basis for 
better understanding of quarkonium production in high energy $p$$+$$p$ 
collisions.

\section*{ACKNOWLEDGMENTS}   

We thank the staff of the Collider-Accelerator and Physics
Departments at Brookhaven National Laboratory and the staff of
the other PHENIX participating institutions for their vital
contributions.  We acknowledge support from the
Office of Nuclear Physics in the
Office of Science of the Department of Energy,
the National Science Foundation,
Abilene Christian University Research Council,
Research Foundation of SUNY, and
Dean of the College of Arts and Sciences, Vanderbilt University
(U.S.A),
Ministry of Education, Culture, Sports, Science, and Technology
and the Japan Society for the Promotion of Science (Japan),
Conselho Nacional de Desenvolvimento Cient\'{\i}fico e
Tecnol{\'o}gico and Funda\c c{\~a}o de Amparo {\`a} Pesquisa do
Estado de S{\~a}o Paulo (Brazil),
Natural Science Foundation of China (People's Republic of China),
Croatian Science Foundation and
Ministry of Science and Education (Croatia),
Ministry of Education, Youth and Sports (Czech Republic),
Centre National de la Recherche Scientifique, Commissariat
{\`a} l'{\'E}nergie Atomique, and Institut National de Physique
Nucl{\'e}aire et de Physique des Particules (France),
Bundesministerium f\"ur Bildung und Forschung, Deutscher
Akademischer Austausch Dienst, and Alexander von Humboldt Stiftung (Germany),
National Science Fund, OTKA, EFOP, and the Ch. Simonyi Fund (Hungary),
Department of Atomic Energy and Department of Science and Technology (India),
Israel Science Foundation (Israel),
Basic Science Research Program through NRF of the Ministry of Education (Korea),
Physics Department, Lahore University of Management Sciences (Pakistan),
Ministry of Education and Science, Russian Academy of Sciences,
Federal Agency of Atomic Energy (Russia),
VR and Wallenberg Foundation (Sweden),
the U.S. Civilian Research and Development Foundation for the
Independent States of the Former Soviet Union,
the Hungarian American Enterprise Scholarship Fund,
and the US-Israel Binational Science Foundation.

\clearpage


%
 
\end{document}